\documentclass{nature}

\usepackage{graphicx}
\usepackage{dcolumn}
\usepackage{bm}
\usepackage{array,multirow}
\usepackage{lineno}
\usepackage{amsmath}
\usepackage{ccaption}

\usepackage{amssymb}
\usepackage[colorlinks=true,citecolor=blue]{hyperref}

\makeatletter
\let\saved@includegraphics\includegraphics
\AtBeginDocument{\let\includegraphics\saved@includegraphics}
\renewenvironment{figure}{\@float{figure}}{\end@float}

\makeatother

\title{Triple-Well Charge Density Wave Transition Driven by Cooperation between Peierls-like Effect and Antiferromagnetic Order in FeGe}
\author{Binhua Zhang$^{1,2}$, Junyi Ji$^{1,2}$, Changsong Xu$^{1,2,\dagger}$,  Hongjun Xiang$^{1,2,3,\dagger}$ }

\begin{document}

\maketitle

\begin{affiliations}
 \item Key Laboratory of Computational Physical Sciences (Ministry of Education), Institute of Computational Physical Sciences, State Key Laboratory of Surface Physics, and Department of Physics, Fudan University, Shanghai 200433, China
 \item Shanghai Qi Zhi Institute, Shanghai 200030, China
 \item Collaborative Innovation Center of Advanced Microstructures, Nanjing 210093, China
\end{affiliations}

\noindent
{B.Z. and J.J.  contributed equally to this work.\\}
{$^{\dagger}$ Corresponding authors: csxu@fudan.edu.cn; hxiang@fudan.edu.cn \\}

\begin{abstract}
Kagome materials provide a promising platform to explore intriguing correlated phenomena including magnetism, charge density wave (CDW), and nontrivial band topology. Recently, a CDW order was observed in antiferromagnetic kagome  metal FeGe, sparking enormous research interests in intertwining physics of CDW and magnetism. Two of the core questions are (i) what are the driving forces of the CDW transition in FeGe and (ii) whether magnetism play a critical role in the transition. Such questions are critical as conventional mechanisms of van Hove singularities and Fermi surface nesting fail to explain the stable pristine phase, as well as the role of magnetism. Here, supported by density functional theory and tight-binding models, we unravel the triple-well CDW energy landscape of FeGe, indicating that both the pristine and CDW phases are locally stable. We point out that an entire downward shift of Ge band, instead of the previously proposed Fe bands, competes with the lattice distortion energy, driving the triple-well CDW transition. It is indeed a cooperation between the Peierls-like effect and the Fermi energy pinning phenomenon, which is distinct from the conventional Peierls effect that drives a double-well transition. Moreover, we demonstrate that the antiferromagnetic order also plays a critical role in driving the CDW transition, through weakening the Fe-Ge hybridization by exchange splitting and lowering the position of Ge-bands with respect to the Fermi energy. Our work thus not only deepens the understanding of the CDW mechanism in FeGe, but also indicates an intertwined connection between the emergent magnetism and CDW in kagome materials.

\end{abstract}

Charge density wave (CDW) is a macroscopic quantum state consisting of electronic charge density modulation accompanied by a periodic lattice distortion, which involves diverse processes such as Fermi surface nesting, Peierls distortion, and electron phonon coupling\cite{gruner1988dynamics, varma1983strong,johannes2008fermi}. The early works on CDW effects, performed with bulk samples of the quasi-one-dimensional metallic crystals, exhibited rich phenomena including  nonlinear electron transport, metal-insulator transitions, and multistable conducting states\cite{zybtsev2010quantized, zaitsev2004finite, wang2013chiral}. Recent years witnessed a rebirth of the field of CDW materials, driven by the emergent kagome materials, e.g., non-magnetic AV$_3$Sb$_5$, where the CDW phases intertwine with superconductivity and band topology\cite{ortiz2019new, ortiz2020cs, ortiz2021superconductivity}. Particularly, as the first magnetic kagome material to manifest CDW, FeGe provides the opportunity for a deeper understanding of the interplay between CDW and magnetism, triggering enormous research interests\cite{miao2022spin,teng2023magnetism,shao2023intertwining, teng2022discovery, yin2022discovery}. 

\begin{figure}[tbp]
  \centering
  \includegraphics[width=16cm]{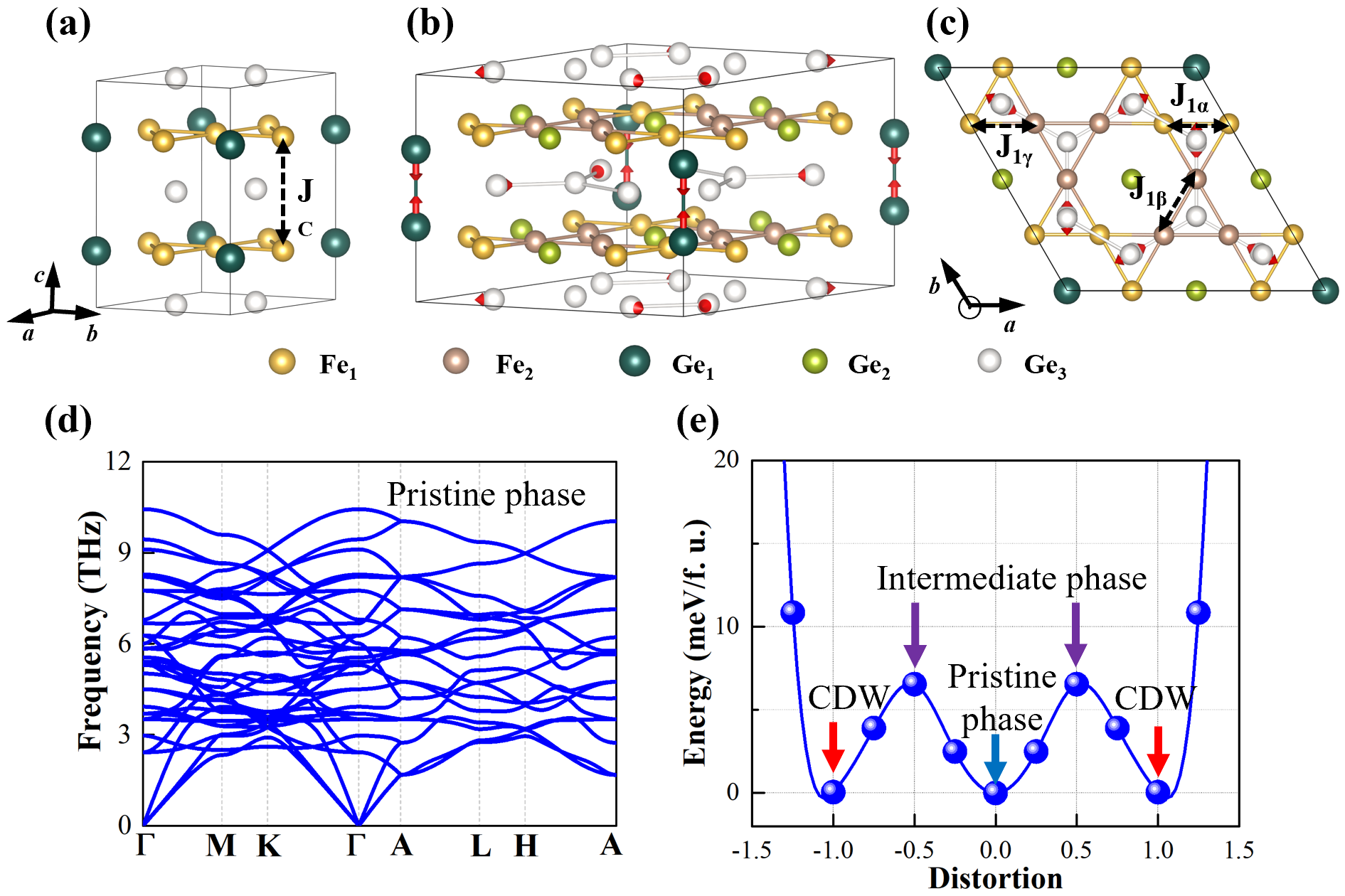}%
  \caption{{\bf Crystal structure, phonon spectrum, and triple-well energy profile of FeGe.} {\bf a} The 1$\times$1$\times$2 pristine crystal structure of FeGe. {\bf b} Three-dimensional view and {\bf c} top view of the 2$\times$2$\times$2 CDW structure. The red arrows show the directions of the atomic displacements in each mode. The exchange paths are denoted by the dashed double arrows. {\bf d} The phonon dispersion of pristine FeGe with the A-AFM order. {\bf e} The triple-well energy profile for the A-AFM FeGe. The distortion stands for the relative amplitude of deformation, with 0.0, $\pm$ 0.5, $\pm$ 1.0 representing the pristine, intermediate, and CDW phases, respectively. The vertical axis stands for the relative total energy with respect to the pristine phase per formula unit (f.u.). Note that the two CDW structures at $\pm$ 1 distortions are equivalent by symmetry.
  }
  \label{fig:structure}
\end{figure}

The high temperature pristine phase of FeGe crystallizes in hexagonal structure with the space group of \emph{P\textup{6}/mmm} (No. 191)\cite{ohoyama1963new}. As depicted in Fig.\ref{fig:structure}a, the Fe atoms form a kagome net interspersed by Ge$_1$ (equivalent to Ge$_2$) atoms and separated by honeycomb layers of Ge$_3$. FeGe exhibits a collinear A-type antiferromagnetic (A-AFM) order below a Néel temperature \emph{T}$_{\rm N}$ $\approx$ 410 K, with spins aligned ferromagnetically within each layer and anti-aligned between adjacent layers\cite{bernhard1984neutron,bernhard1988magnetic}. Below \emph{T}$_{\rm CDW}$ $\approx$ 110 K, the CDW order in FeGe occurs, corresponding to a 2$\times$2$\times$2 modulation of the pristine structure\cite{teng2022discovery}.  
Meanwhile, the magnetic moment enhances at the onset of the CDW order, indicating a strong coupling between the CDW order and A-AFM order\cite{teng2022discovery}. Noteworthily, the pristine phase is actually dynamically stable without exhibiting any imaginary phonon frequencies\cite{shao2023intertwining}, which is distinct from conventional CDW transitions, leaving the driving force elusive.

Despite firm observations, the mechanisms of CDW transition in FeGe remain under hot debate. At first sight, conventional CDW mechanisms bear brunt, attributing the formation of CDW to the presence of  Fe-derived van Hove singularities (VHSs) near Fermi level (\emph E$_\textup F$) induced by the spin splitting\cite{teng2023magnetism}, or the Fermi surface nesting of the kagome electronic states\cite{shao2023intertwining}. However, recent investigations have indicated that (i) the conventional mechanism cannot account for the absence of phonon instabilities in the pristine phase; (ii) only in a small \emph{k}$_z$ range the VHSs are in proximity to \emph E$_\textup F$, which may weaken the nesting functions and contribute scarcely to the formation of CDW\cite{wu2023novel};
(iii) neither signatures of nesting of Fermi surfaces or VHSs nor sizeable electronic energy gaps around the Fermi level have been observed from the latest experiment\cite{zhao2023photoemission}, excluding the dominant role of such conventional mechanisms in driving the CDW phase. Indeed, the very recent experiments and theoretical calculations have begun to associate the CDW phase with the strong dimerization of Ge$_1$-sites along the c-axis\cite{wang2023enhanced,chen2023long}, but the microscopic origin of the triple-well energy landscape and the exact role of magnetism are still unknown. It is therefore highly desirable to reveal the underlying physics of the CDW transition in FeGe.

In this work, we investigate the microscopic mechanism for the formation of the CDW phase by first-principles density-functional theory (DFT) calculations and tight-binding (TB) model analyses. We find that the CDW phase transition in FeGe conforms to a triple-well energy profile, matching well with the phonon stability. In particular, we demonstrate such a unique triple-well CDW transition is driven by a Peierls-like effect, in conjunction with a Fermi energy pinning phenomenon. Specifically, we identify a long-neglected but crucial band consisting of considerable Ge$_1$ \emph{p}$_z$ orbitals. It exhibits a Peierls-like entire downward shift by more than 0.5 eV during the dimerization of Ge$_1$ atoms, and is gradually fully occupied due to the Fermi energy pinning arising from other atoms, thereby lowing the energy. This process, in competition with the lattice distortion energy, leads to the triple-well CDW transition. Furthermore, we reveal the magnetic order plays an indispensable role in driving the CDW transition, through weakening the Fe-Ge hybridization by exchange splitting, as well as lowering the position of Ge-bands with respect to the Fermi energy. Our work thus highlights the cooperation between the Peierls-like effect and magnetic order in such a triple-well CDW system.

\noindent
{\bf Results}

{\bf CDW state of FeGe.} 
Traditionally, the CDW phase is generally constructed from the imaginary phonon modes of the pristine phase. This approach, however, is invalid for FeGe as its pristine phase has no imaginary phonon branch as show in Fig.\ref{fig:structure}d. To determine the possible CDW deformation of FeGe, we adopt different supercells and perform DFT calculations on various configurations, generated from the perturbed genetic algorithm\cite{lou2021pasp} (see Methods). Our results show that the 2$\times$2$\times$2 CDW state, sharing the same space group of \emph{P\textup{6}/mmm} with the pristine phase (Fig.\ref{fig:structure}a), is the most stable among different deformations, consistent with previous studies\cite{shao2023intertwining,chen2023long}. As depicted in Figs.\ref{fig:structure}b,c, in the 2$\times$2$\times$2 CDW state, there are two inequivalent Fe sites and two inequivalent Ge sites in the kagome layer. The adjacent Ge$_1$ atoms of neighboring layers are strongly dimerized via large displacements (0.65 \AA \ per Ge$_1$) along the c-axis\cite{wang2023enhanced}, accompanied by  minor movements (\textless \ 0.08 \AA) of other atoms, including the Kekulé distortion in the Ge$_3$ honeycomb atomic layers\cite{shao2023intertwining}. As a result, the 2$\times$2$\times$2 CDW deformation is mainly characterized by dimerization of Ge$_1$ atoms, distinct from the dominated in-plane displacement of V kagome net found in AV$_3$Sb$_5$\cite{tan2021charge}. 

We then look at the energy evolution during the CDW transition. As shown in Fig.\ref{fig:structure}e, both the pristine and CDW phases are located at the energy minima, consistent with their phonon stabilities [see Fig.\ref{fig:structure}d and Fig.S1 in Supplementary Note (SN) 1\cite{SM}]. The CDW phase shows a comparable energy ($\sim$ 0.05 meV/f.u. higher) with respect to the pristine phase and tends to become the energetically favorable one when taking into account the zero-point energy\cite{shao2023intertwining} or mildly changing of lattice constants induced by different exchange correlation functionals (see SN 6\cite{SM}). The transition from the pristine to CDW phase has to overcome an energy barrier of 6.54 meV/f.u. (that is the energy of the intermediate phase), leading to the peculiar triple-well energy landscape, which is distinct from the double-well scenario in conventional CDW materials.

{\bf Mechanism of the CDW transition.} To reveal the underlying mechanism of the triple-well picture, we track the DFT band structure evolution during the CDW phase transition as shown in Figs.\ref{fig:band}d-f (see SN 2 for detailed band structure evolution and SN 3 for unfolded band structure\cite{SM}). Since the CDW state of FeGe is dominated by the strong dimerization of Ge$_1$ atoms (Fig.\ref{fig:structure}b), we highlight the bands that consist of Ge$_1$ \emph{p}$_z$ orbitals by blue dots, and mark the bands composed of Ge$_2$ \emph{p}$_z$ orbitals by red dots for comparisons. Interestingly, although the Ge$_1$ and Ge$_2$ bands are degenerate in the undistorted pristine phase, they split into two distinct parts in the CDW phase (Figs.\ref{fig:band}d,f). Such a splitting can be understood by the schematic plot in Figs.\ref{fig:band}a,b, where the Ge$_1$ atoms form a triangular sublattice, and the Ge$_2$ atoms form a kagome sublattice. In the CDW phase, the interaction between the two sublattices becomes negligibly weak due to the small overlap of their \emph{p}$_z$ orbitals and large energy difference between Ge$_1$ \emph{p}$_z$ and Ge$_2$ \emph{p}$_z$ orbitals due to the Ge$_1$ dimerization, thereby producing two nearly independent sets of bands. Note that the Ge$_1$ bands presented here all belong to bonding states of the dimerized Ge$_1$ 1D chain (the anti-bonding states lie far above the Fermi energy, see SN 7\cite{SM}), and therefore exhibit a dramatically downward movement after the strong Ge$_1$ dimerization. By contrast, the Ge$_2$ atoms nearly lie in situ during the CDW transition, leading to almost unshifted bands. To further validate such mechanism, we also construct a TB model for the in-plane Ge atoms (see Methods and SN 9\cite{SM}). The calculated TB bands are given in Fig.\ref{fig:band}b, which well captures the band splitting of the two sublattices.

\begin{figure}[tbp]
  \centering
  \includegraphics[width=16cm]{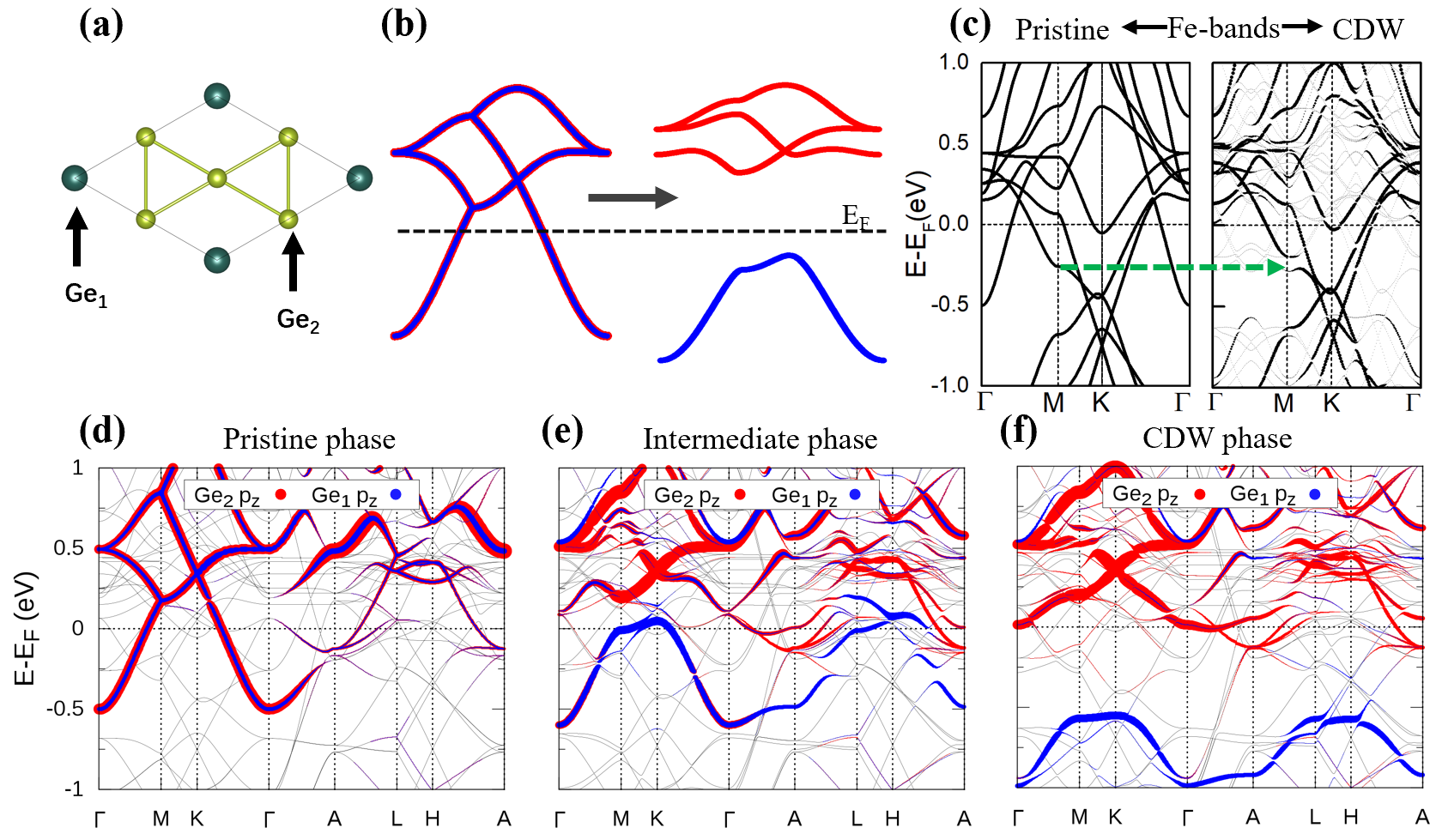}%
  \caption{{\bf  CDW-induced band splitting in FeGe.} {\bf a} The in-plane view of the CDW reconstruction consisting of two types of Ge atoms in kagome layers, where Ge$_1$ with a strong dimerization form a triangular lattice while the rest Ge$_2$ form a kagome lattice. {\bf b} Illustration of CDW-induced band structure modifications from tight-binding model (here, both the nearest-neighbor and next nearest-neighbor interactions are included, see Methods and SN 9\cite{SM} for details). The 2$\times$2$\times$2 pristine phase exhibits a band structure folded from the typical bands of a triangular lattice as shown in the left side. While in the 2$\times$2$\times$2 CDW phase, the degenerate Ge-bands split into two parts: one set of kagome bands above \emph E$_\textup F$ and another triangular band below \emph E$_\textup F$. {\bf c} Unfolded band structures for the pristine phase and the CDW phase, where the green arrow marks the gap opening at the VHS of Fe bands. The calculated band structure for {\bf d} the 2$\times$2$\times$2 pristine phase, {\bf e} the intermediate phase, and {\bf f} the 2$\times$2$\times$2 CDW phase in Fig. \ref{fig:structure}e. The red and blue colors represent the projected bands of Ge$_2$ and Ge$_1$, respectively. Note that the A-AFM order is adopted in the band structure calculations. 
  }
  \label{fig:band}
\end{figure}

\begin{figure}[tbp]
\renewcommand{\figurename}{}
\renewcommand{\thefigure}{}
\DeclareGraphicsExtensions{}
\contcaption \*{Due to the similarity between the band structure evolutions of spin-up and spin-down channels, we only plot the spin-up channel within a FM layer in the A-AFM state here (see the spin-down channel band structure in SN 2\cite{SM}). Note that the plotted Ge-bands in the range of -1 to 1 eV all belong to the bonding states (see SN 7\cite{SM}).\\ }
\end{figure}

We now further understand the triple-well picture by the abovementioned band evolution, in particular the significant downward shift of the Ge$_1$ band during the CDW transition as shown in Figs.\ref{fig:band}d-f. (i) At the beginning of the distortion, both the M-K and L-H paths of Ge$_1$-band sit well above \emph E$_\textup F$. Given that the VHSs of a triangular lattice are located at the M and L points (see SN 4\cite{SM}), the downward shift of Ge$_1$-band results in only a slight decrease in the electronic energy since the VHSs remain unoccupied. However, the distortion leads to a considerable increase in the elastic energy, making the pristine phase be stabilized in an energy minimum. (ii) At the intermediate distortion (1/2 distortion in Fig.\ref{fig:structure}e), the flat M-K and L-H paths of Ge$_1$-band begin to be fully occupied as shown in Fig.\ref{fig:band}e. As a consequence, the downward movement of Ge$_1$ band gives rise to a dramatically decrease in the electronic energy, large enough to overcome the continually increased elastic energy. Accordingly, the total energy starts to drop as shown in Fig.\ref{fig:structure}e. (iii) At fully-developed CDW distortion, the entire Ge$_1$-band has already been fully occupied as shown in Fig.\ref{fig:band}f. Although its downward shift would continue to decrease the electronic energy, the bond length of Ge$_1$-Ge$_1$ dimer becomes so short that further distortion would leads to a sharp increase in the elastic energy arising from the Pauli repulsion. Consequently, the total energy begins to rise again with increasing the distortion, resulting in an energy minimum for the CDW phase. This process, which can also be clearly seen in the partial density of states (see SN 5\cite{SM}), naturally explains the triple-well energy profile of the CDW transition in Fig.\ref{fig:structure}e.

The above band evolution is reminiscent of the well-known Peierls effect\cite{fowler2007electrons}. However, this mechanism is distinct from the conventional Peierls effect. Typically, in the Peierls effect the lower band after splitting is always fully occupied, while in our case it is gradually fully filled due to the Fermi energy pinning phenomenon (The Fermi energy varies from 6.05 to 6.00 eV during the CDW transition according to our DFT calculations, which can be regarded as a fixed value). This is because, although the CDW phase takes electrons from the bands near \emph E$_\textup F$ to fill the Ge$_1$ band, there are plentiful bands contributed by other atoms near \emph E$_\textup F$, which act as an electron reservoir that effectively fix the Fermi energy. It is important to note that without the Fermi energy pinning, the energy profile of CDW transition would be a double-well as in the conventional Perierls effect. Hence, we attribute the unique triple-well picture in FeGe to the cooperation between the Peierls-like effect and the Fermi energy pinning phenomenon.

In addition, we also analyze the band structure reconstruction contributed by Fe atoms, and explain why it is unlikely to drive the CDW transition. As shown in Fig.\ref{fig:band}c, although the CDW phase opens a gap at the VHS (M point as marked by the green dashed arrow), it is located well below the \emph E$_\textup F$, resulting in no energy reduction. Moreover, even if supposing the VHS is located near \emph E$_\textup F$, the related gap opening would likely to trigger a double-well CDW landscape like that in AV$_3$Sb$_5$, rather than the current triple-well transition. Hence, it is the Ge$_1$ bands instead of the previously proposed Fe bands\cite{teng2023magnetism} that drive the triple-well CDW transition in FeGe.

\begin{figure}[tbp]
  \centering
  \includegraphics[width=16cm]{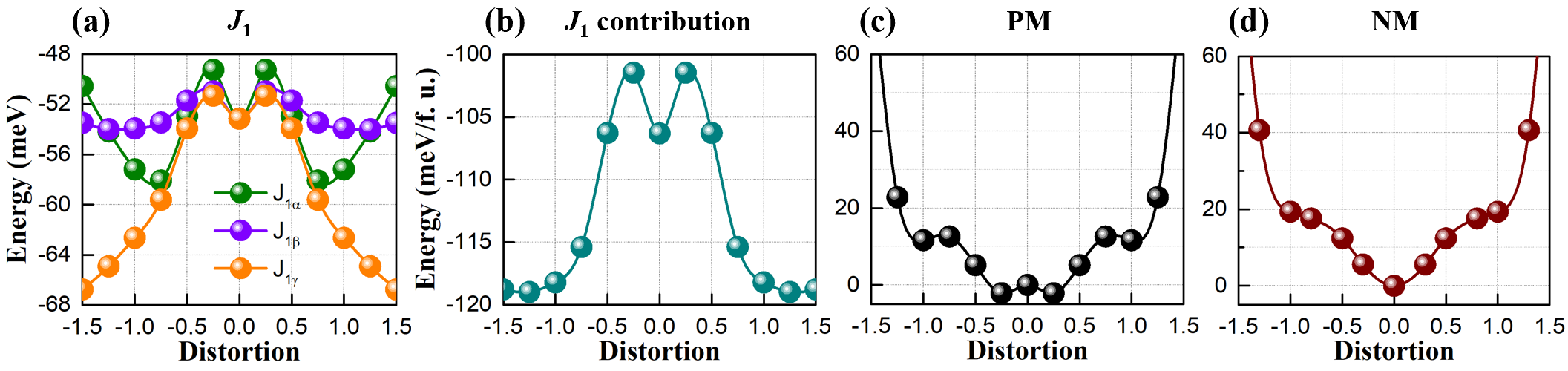}%
  \caption{{\bf  Energy profiles related to magnetism as a function of CDW distortion.} {\bf a} The dependence of magnetic-exchange interaction {\emph J}$_1$ on CDW distortion (using four-state method \cite{xiang2013magnetic} in different distortions). {\bf b} The {\emph J}$_1$ energy contribution ($\sum\limits_{i=\alpha,\beta,\gamma}Z_iJ_{1i}$, where {\emph Z}$_i$ are the multiplication of the paths of {\emph J}$_{1i}$ shown in Fig.\ref{fig:structure}c). {\bf c} The relative total energy with respect to the undistorted pristine phase in paramagnetic (PM) order, which is separated from the total energy by subtracting the mainly magnetic exchange interaction in the A-AFM state. {\bf d} The relative total energy with respect to the undistorted pristine phase in NM order (performing non-spin-polarized DFT calculations). The distortion stands for the relative amplitude of deformation where the $\pm$1 and 0 values are used for CDW and pristine phases, respectively.\\
  }
  \label{fig:magnetism}
\end{figure}

{\bf Energy changes related to magnetism.} Although we have demonstrated that the CDW phase transition in FeGe is primarily driven by the nonmagnetic (NM) Ge atoms, the role of its A-AFM order, which is generally believed to has a strong interplay with its CDW order\cite{teng2022discovery}, is not yet clear. Motivated by this, we calculate the magnetic exchange interactions using the four-state method\cite{xiang2013magnetic}, so as to extract the contribution of magnetism to the CDW transition.
For the undistorted phase, the in-plane first-nearest neighbor yields strong ferromagnetism ({\emph J}$_1$ = -53.16 meV) and dominants over others, while the out-of-plane nearest neighbor favors weak AFM ({\emph J}$_C$ = 8.12 meV) (SN 11\cite{SM}). Consequently, it results in the A-AFM ground-state order with the simulated transition temperature being \emph{T}$_{\rm N}$ $\approx$ 480 K\cite{lou2021pasp}, which is close to the experimental value of \emph{T}$_{\rm N}$ $\approx$ 410 K\cite{bernhard1984neutron,bernhard1988magnetic} (see SN 11 for the simulation\cite{SM}) and is consistent with previous studies\cite{zhou2022magnetic}. 
By comparison, when introducing the CDW structural distortion, the exchange couplings are significantly changed. Here, we take the dominant {\emph J}$_1$ for example. As the two Fe sites are no longer equivalent in the CDW phase, the {\emph J}$_1$ splits into {\emph J}$_{1\alpha}$, {\emph J}$_{1\beta}$, {\emph J}$_{1\gamma}$, as marked in Fig. \ref{fig:structure}c. Their values show a sophisticated evolution as a function of the CDW distortion (Fig.\ref{fig:magnetism}a). Consequently, the {\emph J}$_1$ contribution  (weighted sum of {\emph J}$_{1\alpha}$, {\emph J}$_{1\beta}$ and {\emph J}$_{1\gamma}$) in Fig.\ref{fig:magnetism}b displays an approximately triple-well profile, which lower the energy of CDW phase by $\sim$ -10.67 meV/f.u. compared with the pristine phase, indicating the important role of magnetism in the CDW formation. The magnetic interactions between other nearest neighbors are not discussed here due to their negligible contributions (see SN 11 for detail\cite{SM}). Based on these calculated magnetic interactions, we also separate the corresponding paramagnetic (PM) contribution from the total energy as depicted in Fig. \ref{fig:magnetism}c. The increasing trend from the pristine phase to the CDW phase, that is, the energy cost for structural distortions, provides indirect evidence for the interplay between the CDW and magnetism. Besides, we also plot the energy profile for FeGe in NM state (Fig.\ref{fig:magnetism}d, see SN 8 for analysis on NM state\cite {SM}), which is similar to that of PM state in Fig.\ref{fig:magnetism}c, again confirming that A-AFM order is crucial for the CDW formation.

\begin{figure}[tbp]
  \centering
 \includegraphics[width=16cm]{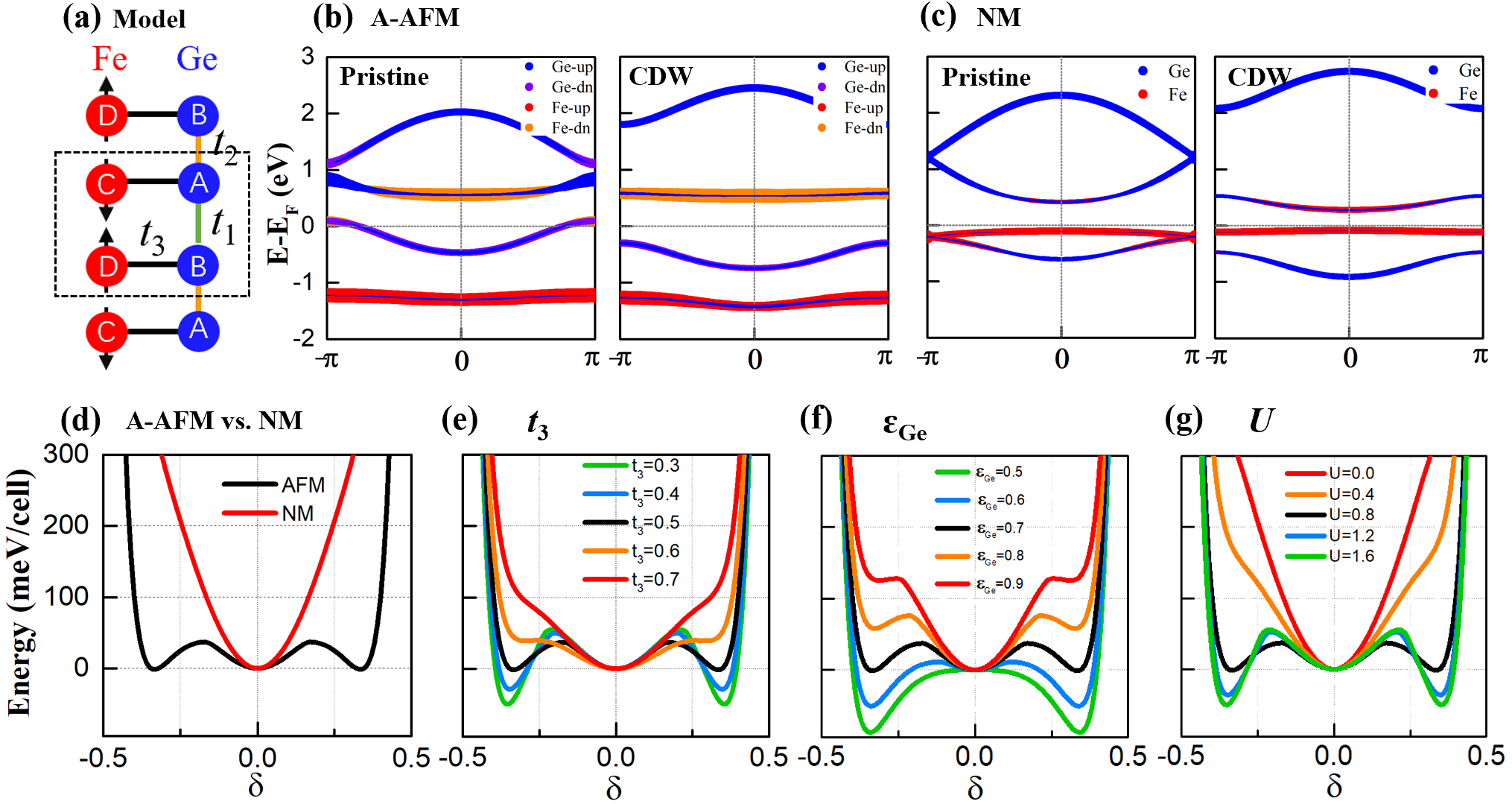}%
  \caption{{\bf Schematic diagrams of tight-binding calculations of FeGe.} {\bf a} Illustration of the one-dimensional atomic chain model along $z$-axis. The black arrows show the A-AFM order. {\bf b, c} Calculated TB band structures of the pristine and CDW states for FeGe in the A-AFM and NM order. The fitted TB parameters can be found in the main text. {\bf d} Relative total energies with respect
to the undistorted pristine phase as a function of CDW distortion. {\bf e-g} The dependence of energy profiles on the Fe-Ge hybridization \emph{t}$_3$, on-site energy $\varepsilon$$_{\textup{Ge}}$, and exchange splitting \emph{U}. Here, the vertical axis stands for the relative total energy with respect to the undistorted pristine phase (in unit of meV per cell in black dashed line in {\bf a}). The horizontal axis $\delta$ stands for the relative deformation with respect to the undistorted Ge-Ge bond. The black lines represent the energy curves taking the initial set values.} 
  \label{fig:TB}
\end{figure}

{\bf The interplay of magnetism and CDW revealed by tight-binding model.} 
Tight-binding model is developed to get further insights into the interplay of magnetism and CDW transition. As shown in Fig.\ref{fig:TB}a, a simplified one-dimensional atomic structure is adopted with  Ge chain being along $z$-axis and each Ge being connected with one Fe.
The corresponding Hamiltonian then reads
\begin{equation}
\mathcal{H}(\bm k)= \begin{pmatrix}
   \mathcal{H}_\uparrow & 0 \\
    0 & \mathcal{H}_\downarrow
    \end{pmatrix},
    \label{Eq:1}
\end{equation}
where
\begin{equation}
\mathcal{H}_{\uparrow/\downarrow} (k)=\emph{E}_\textup F + \begin{pmatrix}
   \varepsilon_{\textup Ge} & \emph{t}_1+\emph{t}_2e^{-ik} & \emph{t}_3 & 0 \\
    \emph{t}_1+\emph{t}_2e^{ik} & \varepsilon_{\textup Ge} & 0 & \emph{t}_3 \\
    \emph{t}_3 & 0 & \varepsilon_{\textup Fe} \mp U & 0 \\
    0 & \emph{t}_3 & 0 & \varepsilon_{\textup Fe} \pm U
    \end{pmatrix},
    \label{Eq:2}
\end{equation}
\emph{E}$_\textup F$ is the Fermi energy; $\varepsilon$$_{\textup Ge}$ and $\varepsilon$$_{\textup Fe}$ are the on-site energy of Ge-\emph{p}$_z$ and Fe-\emph{d} orbitals with respect to \emph{E}$_\textup F$, respectively; \emph{U} characterizes the exchange splitting (2\emph{U}=$\mid$$\varepsilon_{Fe,\uparrow}$-$\varepsilon_{Fe,\downarrow}$$\mid$, \emph{U} = 0 for NM case); \emph{t}$_i$ \{\emph{i} = 1, 2, 3\} represent the hopping integral between different ions as labeled in Fig. \ref{fig:TB}a. According to Harrison’s method \cite{harrison2004elementary}, the dependence of \emph{t}$_i$ on the bond distortion can be written as \emph{t}$_1$= $\frac{\emph{t}_0}{(1+\delta)^{2}}$ and \emph{t}$_2$= $\frac{\emph{t}_0}{(1-\delta)^{2}}$ (\emph{t}$_3$ is assumed to be a constant due to the very slightly change of the Fe-Ge bond), where \emph{t}$_0$ is the hopping integral in the pristine phase, $\delta$ stands for the relative deformation with respect to the pristine undistorted Ge-Ge bond, with $\delta$ = 0.00, 0.16, 0.32 corresponding to the pristine, intermediate and CDW phases in Fig. \ref{fig:structure}e, respectively.

After diagonalizing the Hamiltonian, the occupation energy of the system can be determined by integrating the eigenvalues below the \emph E$_\textup F$. Here, we take into account the aforementioned Fermi energy pinning phenomenon, which arises from the other atoms that are not included in the TB model. Specifically, there exist plentiful bands near the Fermi energy in the DFT electronic structures as shown in Figs.\ref{fig:structure}d-f, which act as a reservoir of electrons that effectively pins the Fermi energy, regardless of the change of occupation number of Ge$_1$ bands during the phase transition. Thus, the almost invariant Fermi energy \emph E$_\textup F$, instead of the occupation number, is fixed in our TB model. Considering this, as well as the lattice distortion energy, the total energy can then be expressed as
\begin{equation}
\emph{E}_{\textup {tot}}={\sum\limits_{i=1, 8}\int\epsilon(i,k)f(i,k)dk}-N_{\textup{occ}}E_{\textup{F}}+\emph{B}[\frac{1}{(1+\delta)^{12}}+\frac{1}{(1-\delta)^{12}}]+\emph{B}_2\delta^2.
    \label{Eq:3}
\end{equation}
where the first term is the integrated occupation energy of all eigenvalues $\epsilon(i,k)$ with occupations $f(i,k)$; the second term gathers the energy contribution from the electrons transfer between the TB bands and the reservoir bands at the \emph E$_\textup F$, with \emph{N}$_{\textup {occ}}$=$\sum\limits_{i=1, 8}\int f(i,k)dk$ the integrated number of occupied electrons in the TB bands; the third term represents the lattice distortion energy of Ge-Ge interactions, extracted from the repulsion part of the Lennard–Jones potential\cite{wei2021lead}; and the last term denotes the Fe-Ge interactions, which is approximated by a simple harmonic form due to the slight change in the Fe-Ge distance from the equilibrium position. By fitting to the DFT electronic structures and total energies (see SN 15\cite{SM} for detail), we obtain \emph{E}$_\textup F$ = 6.0 eV, $\varepsilon$$_{\textup{Ge}}$ = 0.7 eV, $\varepsilon$$_{\textup{Fe}}$ = -0.3 eV, \emph{U} = 0.8 eV, \emph{t}$_0$ = 0.6 eV, \emph{t}$_3$ = 0.5 eV, \emph{B} = 0.001 eV, \emph{B}$_2$ = 6.8 eV for the A-AFM FeGe. While for the NM FeGe (\emph{U} = 0 eV), although the absolute values of \emph{E}$_\textup F$+$\varepsilon$$_{\textup{Ge}}$ and \emph{E}$_\textup F$+$\varepsilon$$_{\textup{Fe}}$ remain the same, the \emph{E}$_\textup F$ changes to 5.7 eV (see SN 9\cite{SM}), which effectively leads to $\varepsilon$$_{\textup{Ge}}$ = 1.0 eV, $\varepsilon$$_{\textup{Fe}}$ = 0.0 eV, consistent with the Stoner model that the $\varepsilon$$_{\textup{Fe}}$ lies near the Fermi energy.

The band structures based on the TB model with the fitted parameters are plotted in Figs. \ref{fig:TB}b,c. A shift of the entire Ge-band is clearly captured during the CDW phase transition for both A-AFM and NM orders, in line with the DFT calculations (see SN 9\cite{SM}). The relative total energy as a function of the distortion $\delta$ is further displayed in Fig.\ref{fig:TB}d, which well confirms the triple-well profile in the A-AFM order and the single-well profile in the NM order. Such distinct energy profiles are mainly associated with the differences in the band structures between the A-AFM and NM states: (i) in the A-AFM state, the Fe-Ge hybridization near \emph E$_\textup F$ is significantly weaker, since the Fe orbitals are pushed far away from the \emph E$_\textup F$ by the exchange splitting \emph{U}; (ii) the location of Ge-band with respect to \emph E$_\textup F$ (characterized by $\varepsilon$$_{\textup{Ge}}$) in the pristine state is higher in the NM order than in the A-AFM order, which makes it more difficult for the transition to happen. Such TB results are in line with the DFT calculations (see SN 9\cite{SM}). 

 We now further understand the effect of magnetism on the CDW formation by adjusting the TB parameters independently. We first modify the strength of Fe-Ge hybridization by directly tuning the parameter \emph{t}$_3$. As shown in Fig.\ref{fig:TB}e, the energy profile evolves from a single-well to a triple-well as \emph{t}$_3$ decreases. Alternatively, the Fe-Ge hybridization can also be indirectly adjusted via modifying the exchange splitting \emph{U}. For \emph{U} = 0 (NM order), the Fe orbitals are close to Ge orbitals near \emph E$_\textup F$, giving rise to strong Fe-Ge hybridization. As the value of \emph{U} increases, the Fe-band is pushed away from \emph E$_\textup F$, which weakens the Fe-Ge hybridization, and the energy profile gradually changes from a single-well to a triple-well as shown in Fig.\ref{fig:TB}g. Such results with varying $t_3$ and $U$ indicate that a weaker Fe-Ge hybridization is beneficial to form the CDW state with a triple-well potential. Based on finding, we infer that a tensile strain is likely to facilitate the formation of the CDW states due to the weakening of Fe-Ge hybridization under the lattice expansion, which is indeed verified by our further DFT calculations (see SN 6\cite{SM}). On the other hand, the initial location of Ge-band ($\varepsilon$$_{\textup{Ge}}$) is also responsible for the formation of the CDW state (Fig.\ref{fig:TB}f): (i) for a large $\varepsilon$$_{\textup{Ge}}$, it may cost a high energy to overcome the lattice distortion before the occupation of Ge-bands, causing a single-well energy profile; (ii) for a small $\varepsilon$$_{\textup{Ge}}$ (near \emph E$_\textup F$), it would lower energy to form the CDW state by the instant splitting of Ge-bands, leading to a double-well energy profile; (iii) for a intermediate $\varepsilon$$_{\textup{Ge}}$, a triple-well profile appears, in line with the aforementioned DFT description. With such further information from TB, the mechanisms is clear for not only the CDW transition of FeGe, but also the fact that the isomorphic FeSn show no experimentally observed CDW phase transition, which probably attributes to the larger “$\varepsilon$$_{\textup{Ge}}$” in the case of FeSn (see SN 10 for details\cite{SM}). 
Given all of the above discussion, it is now clear that the A-AFM order, is critical for forming the triple-well CDW phase transition, since it weakens the Fe-Ge hybridization through the exchange splitting \emph{U}, and possesses a lower position of Ge-bands with respect to the Fermi energy.

\noindent
{\bf Conclusions}

In summary, by applications of DFT and TB model analyses, the mechanism of triple-well CDW in FeGe is revealed as a cooperation between the Peierls-like effect and magnetism. The entire band of Ge$_1$ atom, which hosts a large dimerization, is found to be responsible for the formation of CDW phase. Moreover, it is found that the A-AFM order is critical for lowering energy and stabilizing the CDW phase, through interplay with the structural distortion. 
We expect the presently discovered new mechanisms not only can be used to understand the CDW transition in FeGe, but also shed light on other systems with intertwined magnetism and CDW.



\clearpage
\begin{methods}
{\bf DFT calculations}\\
DFT calculations are performed using the projector-augmented wave method\cite{blochl1994projector} as implemented in the Vienna\emph{ab initio} Simulation Package (VASP)\cite{kresse1996efficiency, kresse1996efficient2}. The exchange-correlation interactions are described by the generalized-gradient approximation with the Perdew-Burke-Ernzerhof (PBE) functional\cite{perdew1996generalized}. An energy cutoff of 350 eV is used for the plane wave basis expansion. All atomic positions are fully relaxed until the force acting on each atom became smaller than $1\times10^{-6}$ eV/\AA. To determine the possible CDW deformation of FeGe, we adopt the genetic algorithm for structure research, as implemented in the PASP software\cite{lou2021pasp}. In the 1$\times$1$\times$2 pristine phases, the calculated lattice constants are 4.97 \AA (\emph{a} = \emph{b}) and 8.11 \AA (\emph{c}). The Brillouin zone is sampled with a 16$\times$16$\times$10 k-point mesh in the 1$\times$1$\times$2 primitive cell. A 8$\times$8$\times$10 mesh is used for the 2$\times$2$\times$2 supercell. k-meshes of 28$\times$28$\times$18 and 16$\times$16$\times$18 are used to calculate the band structures of the pristine and 2$\times$2$\times$2 CDW phases, respectively. To visualize the effect of structural deformation to the electronic structure, the effective band structure of CDW states is unfolded to the pristine Brillouin zone by the band-unfolding method using vaspkit code\cite{wang2021vaspkit}. The phonon dispersion is calculated by using the finite displacement method as implemented in the phonopy code\cite{togo2015first}. In calculating the phonon spectrum, a 9$\times$9$\times$6 mesh is used for 2$\times$2$\times$2 supercell of the pristine phase, and 8$\times$8$\times$10 mesh for 1$\times$1$\times$1 supercell of the CDW phases. The four-state energy mapping method\cite{xiang2013magnetic} was performed to obtain magnetic parameters from DFT total energies, for which the $\sqrt{3}$$\times$$\sqrt{3}$$\times$1 CDW supercell and 4$\times$4$\times$10 mesh were adopted. In calculating single-ion anisotropy, the spin-orbit coupling was included. Monte Carlo (MC) simulations were performed using the calculated magnetic exchange interactions. The 30$\times$30$\times$1 supercells were adopted in the study. For each configuration, 10000 and 100000 MC steps per site were performed for equilibrating the system and statistical averaging, respectively.
\\
{\bf TB models}\\
In the one-dimensional atomic chain model, the single electron Schrödinger equation can be repressed as
\begin{equation}
H\Psi_{\bm k} (\bm r)=\epsilon_{\bm k}\Psi_{\bm k} (\bm r).
    \label{Eq:4}
\end{equation}
where the wave function and Hamiltonian can be written as
\begin{equation}
\Psi_{\bm k} (\bm r)=\sum_{u=1}^4c_u\phi_{\bm {u,k}}(\bm r),\ \phi_{\bm {u,k}}(\bm r)=\frac{1}{\sqrt{N_c}}\sum_Le^{i{\bm {kR_L}}}\phi_{\bm u}(\bm {r-R_L-\tau_u}),
    \label{Eq:5}
\end{equation}
\begin{equation}
H_{vu}({\bm k})=\sum_Me^{i{\bm {kR_M}}}\left\langle{\phi_v(\bm {r-\tau_v})|H|\phi_u(\bm {r-\tau_u-R_M})}\right\rangle,
    \label{Eq:7}
\end{equation}
where \emph{c}$_u$ is the weights of the linear combination, \emph{${\bm R}$}$_L$ is the lattice vector, \emph{${\bm \tau}$}$_u$ describes the real space translation vector of \emph{u}-atom in the unit cell (see Fig.\ref{fig:TB}a, here, \emph{u}=1,2,3,4, denoting Ge$_A$, Ge$_B$, Fe$_C$, Fe$_D$, respectively). \emph{H}$_{vu}$(${\bm k}$) from Eq.\ref{Eq:7} are the matrix elements of the Hamiltonian, which can be further expressed using the on-site energy $\varepsilon$$_i$ \{\emph{i} = Ge, Fe\} and hopping terms \emph{t}$_i$ \{\emph{i} = 1, 2, 3\} (see Fig.\ref{fig:TB}a). Then the tight-binding Hamiltonian is given as Eq.\ref{Eq:2}.

Likewise, the Hamiltonian for the spinless Ge atoms in two-dimensional lattice can be expressed as
\begin{equation}
H=\sum_{i}\varepsilon_i\hat{a}_i^+\hat{a}_i+\sum_{i,j}t_{i,j}\hat{a}_i^+\hat{a}_j,
    \label{Eq:9}
\end{equation}
where $\varepsilon$$_i$ is the on-site energy of atom \emph{i} (in the 2×2 cell, there are four Ge atoms, denoting 1, 2, 3, 4, respectively, see Fig.S15\cite{SM}). \emph{t}$_{ij}$ is the hopping term between \emph{i} and \emph{j}, belonging to different sublattices. Corresponding Hamiltonian \emph{H} can be expressed as
\begin{equation}
\mathcal{H}_2(k)= \begin{pmatrix}
   \varepsilon_1 & \emph{t}_{12}\sum_{\Vec{r}_{12}}e^{-i\Vec{k}\cdot\Vec{r}_{12}} & \emph{t}_{12}\sum_{\Vec{r}_{13}}e^{-i\Vec{k}\cdot\Vec{r}_{13}} & \emph{t}_{12}\sum_{\Vec{r}_{14}}e^{-i\Vec{k}\cdot\Vec{r}_{14}} \\
    \emph{t}_{12}\sum_{\Vec{r}_{12}}e^{i\Vec{k}\cdot\Vec{r}_{12}} & \varepsilon_2 & \emph{t}_{34}\sum_{\Vec{r}_{23}}e^{-i\Vec{k}\cdot\Vec{r}_{23}} & \emph{t}_{34}\sum_{\Vec{r}_{24}}e^{-i\Vec{k}\cdot\Vec{r}_{24}} \\
    \emph{t}_{12}\sum_{\Vec{r}_{13}}e^{i\Vec{k}\cdot\Vec{r}_{13}} & \emph{t}_{34}\sum_{\Vec{r}_{23}}e^{i\Vec{k}\cdot\Vec{r}_{23}} & \varepsilon_2 & \emph{t}_{34}\sum_{\Vec{r}_{34}}e^{i\Vec{k}\cdot\Vec{r}_{34}} \\
    \emph{t}_{12}\sum_{\Vec{r}_{14}}e^{i\Vec{k}\cdot\Vec{r}_{14}} & \emph{t}_{34}\sum_{\Vec{r}_{24}}e^{i\Vec{k}\cdot\Vec{r}_{24}} & \emph{t}_{34}\sum_{\Vec{r}_{34}}e^{i\Vec{k}\cdot\Vec{r}_{34}} & \varepsilon_2
    \end{pmatrix}.
    \label{Eq:10}
\end{equation}
Here, $\Vec{r}$$_{ij}$ represent the possible combination vectors between \emph{i} and \emph{j}, which can contain both the nearest-neighbor and next nearest-neighbor interactions. After solving the Hamiltonian matrix, we can have: (i) for the undistorted pristine phase, all four Ge atoms are equivalent, $\varepsilon$$_1$ = $\varepsilon$$_2$, \emph{t}$_{12}$ = \emph{t}$_{34}$, corresponding to the degenerate energy eigenvalues as plotted in Fig.\ref{fig:band}b (left side); (ii) when the CDW distortion occurs, the initial equivalent Ge atoms split into one triangular band consisting of Ge$_1$ atom and three kagome bands consisting of Ge$_2$ atoms (right side in Fig.\ref{fig:band}b), revealing good agreement with DFT calculations.


\end{methods}


\begin{addendum}
 \item[Data Availability] All the data supporting the findings of this study are provided within this paper and its 
Supplementary Information files. All the raw data generated in this study are available from the corresponding author upon reasonable request. 
 \item We thank Prof. Guoqing Chang for useful discussions. We acknowledge financial support from the Ministry of Science and Technology of the People's Republic of China  (No. 2022YFA1402901), the support from NSFC (grants No. 11825403, 11991061, 12188101, 12174060, and 12274082), and the Guangdong Major Project of the Basic and Applied Basic Research (Future functional materials under extreme conditions--2021B0301030005). 
C. X. also acknowledge the support from Shanghai Science and Technology Committee (grant No. 23ZR1406600).
B. Z. also acknowledge the support from China Postdoctoral Science Foundation (grant No. 2022M720816).
 \item[Correspondence] Correspondence should be addressed to C.X. (csxu@fudan.edu.cn) and H.X. (hxiang@fudan.edu.cn).
\end{addendum}

\bibliographystyle{naturemag}
\noindent
{\bf References}
\bibliography{FeGe.bib}


\end{document}